\documentclass[11pt,twoside]{article}

\usepackage{asp2006}
\usepackage{epsf}
\usepackage[dvips]{graphicx}
\usepackage{lscape}

\begin{document}
\title{The Progress of Solar Cycle 24 at High Latitudes}
\author{Richard C. Altrock}
\affil{Air Force Research Laboratory, NSO/SP, PO Box 62, Sunspot, NM 88349, USA}

\begin{abstract} 
The ``extended'' solar cycle 24 began in 1999 near 70$\deg$ latitude,
similarly to cycle 23 in 1989 and cycle 22 in 1979.  The extended cycle
is manifested by persistent Fe XIV coronal emission appearing near 70$\deg$
latitude and slowly migrating towards the equator, merging with
the latitudes of sunspots and active regions (the ``butterfly diagram'')
after several years.  Cycle 24 began its migration at a rate 40\% slower
than the previous two solar cycles, thus indicating the possibility of
a peculiar cycle.  However, the onset of the ``Rush to the Poles'' of
polar crown prominences and their associated coronal emission, which
has been a precursor to solar maximum in recent cycles (cf. Altrock
2003), has just been identified in the northern hemisphere.
Peculiarly, this ``Rush'' is leisurely, at only 50\% of the rate in the
previous two cycles.  The properties of the current ``Rush to the Poles''
yields an estimate of 2013 or 2014 for solar maximum.
\end{abstract}

%%% MAIN BODY OF TEXT GOES HERE. CONSULT "INSTRUCTIONS FOR AUTHORS USING
%%% LATEX2E MARKUP", SECTIONS 2.3-2.6 FOR HELP WITH EQUATIONS, FIGURES,
%%% AND TABLES.
%\subsection{}   %%% Second level section head (remove "%" symbol)
%\subsubsection{}   %%% Lowest level section head (remove "%" symbol)
%\section*{}    %%% Unnumbered top level section head (remove "%" symbol)
%\subsection*{}   %%% Unnumbered second level section head (remove "%" symbol)

\section{Introduction}   

Altrock (1997) and earlier authors (cf. Wilson et al. 1988) discussed
the high-latitude ``extended'' solar cycle seen in the Fe XIV corona
prior to the appearance of sunspots and active regions at lower
latitudes. For example, persistent coronal emission appeared near
70$\deg$ latitude in 1979 and 1989 and slowly migrated towards the
equator, merging with the latitudes of sunspots and active regions
after several years.  Wilson et al. (1988) discussed other
observational parameters that have similar properties, and this was
updated by Altrock, Howe and Ulrich (2008) for torsional oscillations.

Altrock (2007) showed that the high-latitude coronal emission was
situated above the high-latitude neutral line of the large-scale
photospheric magnetic field seen in Wilcox Solar Observatory synoptic
maps, thus implying a connection with the solar dynamo.

Altrock (2003) discussed coronal emission features seen in Fe XIV
which, prior to solar maximum in cycles 21 - 23, appeared above
50$\deg$ latitude and began to move towards the poles at a rate of 8 to
11 $\deg$$yr^{-1}$. This motion was maintained for a period of 3 or 4
years, at which time the emission features disappeared near the poles.
This phenomenon has been referred to as the ``Rush to the Poles''
(RttP).  It was first identified in solar-crown prominences, and it was
first observed in the corona by Waldmeier (1964).  Altrock concluded
that (i) the maximum of solar activity, as defined by the smoothed
sunspot number, occurred 1.5 $\pm$ 0.2 yr before the extrapolated
linear fit to the RttP reached the poles, and (ii) the RttP could be
used to predict the date of solar-cycle maximum up to three years prior to its
occurrence.  He stated that, ``For solar cycle 24, a prediction of the
date of solar maximum can be made when the RttP becomes apparent,
approximately eleven years after its cycle-23 onset on 1997.58, or 2008 -
2009. When that occurs, the average slope for cycles 21 - 23, 9.38
$\pm$ 1.71 $\deg$$yr^{-1}$, can be used to predict the arrival date of
the RttP at the poles, and then the average lag [time between solar
maximum and the date the extrapolated linear fit to the RttP reached
the poles], 1.52 $\pm$ 0.20 yr, can be used to predict the date of
solar maximum ...''

\section{Observations}   

\begin{figure}[!ht]
\plotone{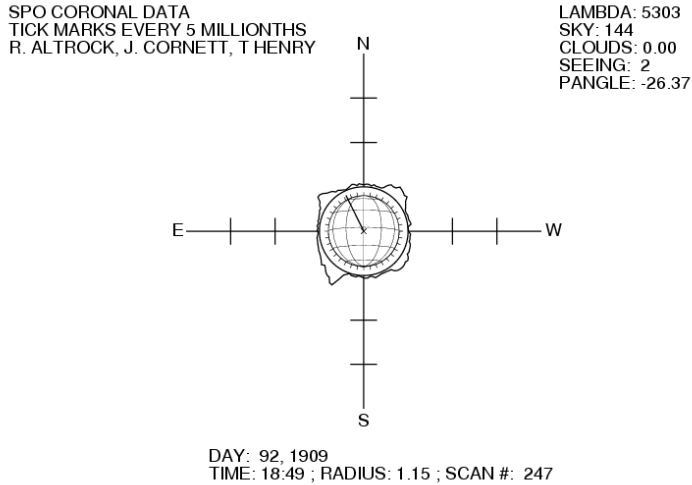}
\caption{\label{fig:fig1} Sample polar plot of Fe XIV intensity at 1.15 Ro at solar minimum.  Intensity is zero at outer circle.}
\end{figure}

Observations of the Fe XIV 530.3 nm solar corona have been attempted
three to seven times a week since 1973 with the photoelectric coronal
photometer and 40-cm coronagraph at the John W. Evans Solar Facility of
the National Solar Observatory at Sacramento Peak (Fisher 1973 and
1974; Smartt 1982).  The photometer automatically removes the
highly-variable sky background.  Scans at 0.15 solar radii
($R_{\odot}$) above the limb every 3$\deg$ in position angle show
coronal features overlying active regions, prominences, large-scale
magnetic field boundaries, etc.  Observations near solar minimum
continue to show coronal emission overlying high-latitude neutral lines
even when there are no active regions at the limb.  
Figure 1 shows a sample solar-minimum scan.  Note (i) 
a lack of low-latitude active region emission and (ii) emission
occurring at higher latitudes.

\begin{figure}[!ht]
\plotone{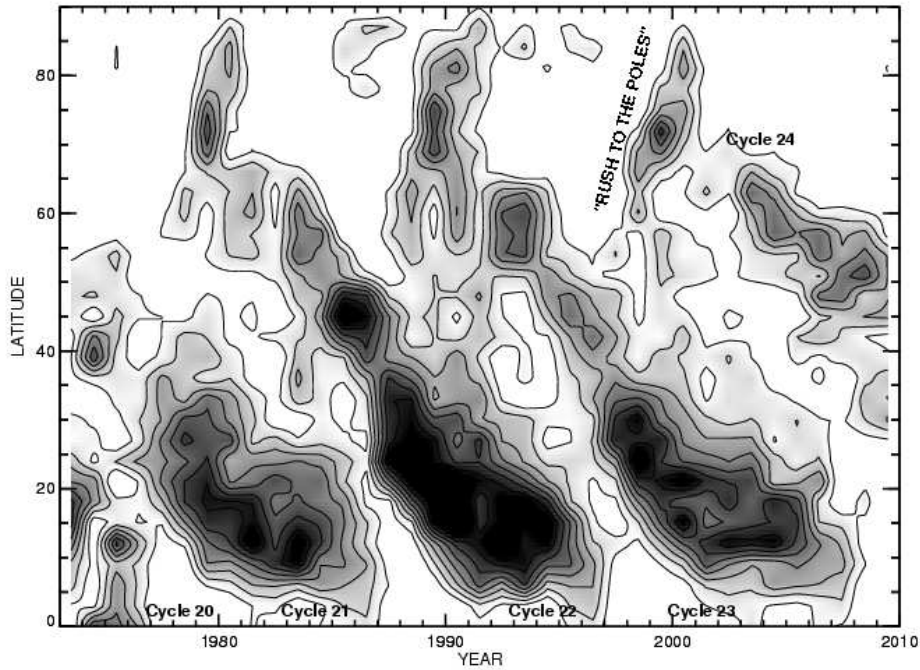}
\caption{\label{fig:fig2} 
Annual northern plus southern hemisphere averages of the number of Fe
XIV intensity maxima from 1973 through 2009.  The ``Rush to the Poles''
around 2000 is indicated, as well as the extended solar cycle 24,
beginning in approximately 1999.}
\end{figure}

\section{Procedure}   

As discussed in Altrock (1997), the daily scans of the corona in Fe XIV
at 1.15 $R_{\odot}$ are examined to determine the location in latitude
of local intensity maxima, and each maximum is plotted on a synoptic
map of latitude vs. time.  Altrock (1997) Figure 3 shows such a
synoptic map from 1973 to 1996.  Note that nowhere in this analysis is
the value of the intensity used, and that allows tracking of very faint
features.  To clarify the solar-cycle behavior of the intensity maxima,
the number of points at each latitude in the synoptic map is averaged
over a given time interval.  This process allows us to correct the
figure for days of missing data, which is an important step in order to
correctly interpret the data.  Figure 2 shows annual
averages of the number of intensity maxima, also averaged over the
north and south hemispheres.

\section{Discussion}   

In Figure 2 we can clearly see the nature of extended solar cycles and
RttP over the last 30+ years.  Extended solar cycles begin near 70$\deg$
latitude and end near the equator about 18 years later, as can be seen
in cycles 22 and 23.  Note that cycle 24 began similarly to cycles 22
and 23; however, it has been migrating equatorward more slowly.  The
rates for cycles 22 - 24 have been -5.3, -4.7 and -3.1 $\deg$$yr^{-1}$,
respectively.  Most recently, emission took a sudden jump down to
around 30$\deg$ , and there is an indication that the RttP could be
developing.  This suggests to use a higher-resolution (if noisier) graphic.

\begin{figure}[!ht]
\plotone{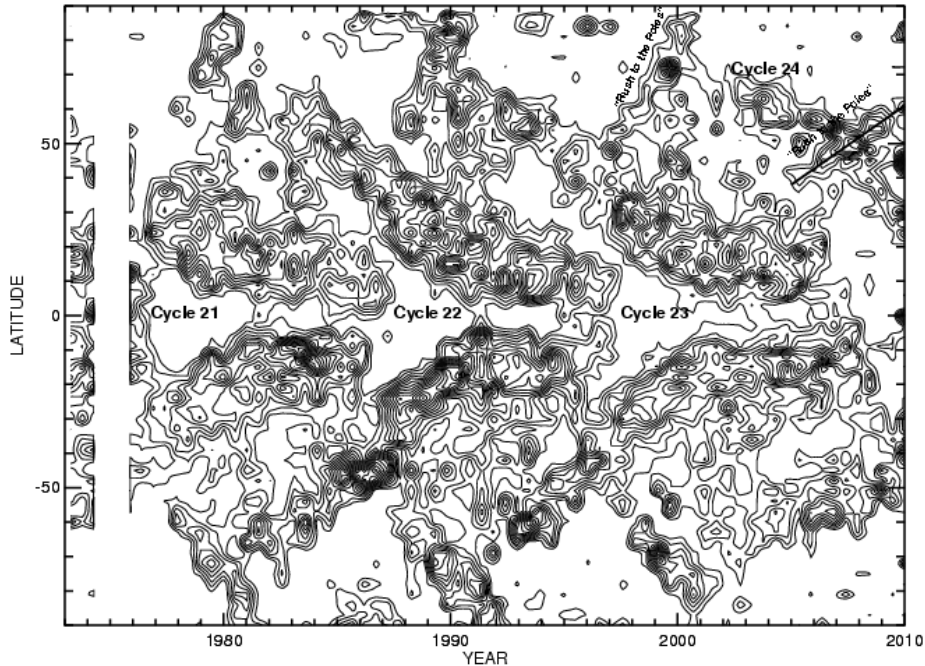}
\caption{\label{fig:fig3} 
Seven-rotation (approximately semiannual) averages of the number of Fe
XIV emission maxima from 1973 through 2009.  Note the cycle 24 ``Rush to
the Poles'' indicated by the label and linear fit in the upper right corner.}
\end{figure}

Figure 3 shows the data with a temporal average of 189
days, or seven 27-day rotation periods (approximately semiannual).  In
addition we now examine both hemispheres independently.  Here we see
that the extended cycle 24 (the equatorward-moving emission) appears to
have recently split into two branches, most easily seen in the northern
hemisphere.  This has occurred previously, notably in the northern
hemisphere in approximately 1991, after which the lower latitude branch
eventually disappeared.  So it is difficult to say to what the current
split may lead.

The more interesting development is the appearance in 2005 of the RttP
in the northern hemisphere, marked by a label, ``Rush to the Poles'', and
a linear fit, both seen in the upper right-hand corner of the graph.
No such feature is yet evident in the southern hemisphere, which only
may mean it is not yet visible in these noisy data or that it is
delayed.  In any case, we can use the northern hemisphere data as an
indicator of when solar maximum will occur.  The current RttP rate is
estimated to be 4.6$\deg$$yr^{-1}$ (recall 9.4 $\pm$ 1.7 $\deg$$yr^{-1}$ 
average in the previous three cycles [Altrock 2003]).  This 50\% lower
rate makes the earlier suggestion to use the previous higher rate to
estimate the time of cycle maximum invalid (see discussion in
Introduction).

At the current rate, the extrapolated RttP will reach the north pole at
2016.3.  If we apply the previously-determined 1.5 $\pm$ 0.2 yr offset
between solar maximum and arrival at the poles (see Introduction), this
would imply solar maximum at 2014.8 $\pm$ 0.2.  However, using that offset
could be somewhat dubious, considering the slow ``rush'' this cycle.

A method that is possibly more reliable is to use the property that
solar maximum occurs when the center line of the RttP reaches a
critical latitude.  In the previous three cycles this latitude was
76$\deg$, 74$\deg$ and 78$\deg$, for an average of 76$\deg$ $\pm$
2$\deg$ [this can be determined from the figures in Altrock (2003)].
At the current rate, this will occur at 2013.3 $\pm$ 0.5.  If the RttP
rate increases, solar maximum would be earlier, although there is no
reason to believe that this will occur.

Thus, the two methods using the coronal ``rush to the poles'' result in
predictions for solar maximum at 2013.3 $\pm$ 0.5 and 2014.8 $\pm$ 0.5, or
2013-2014.

Nothing in this analysis yields the sunspot number to be expected at
solar maximum.

\section{Conclusions}   

The location of Fe XIV intensity maxima in time-latitude space displays
an 18-year progression from near 70$\deg$ to the equator, which has been
referred to as the ``extended'' solar cycle.  Cycle 24 emission began
proceeding towards the equator similarly to previous cycles, although
at a 40\% slower rate.  In addition, in 2009 the northern hemisphere
``Rush to the Poles'' became evident and is proceeding at a 50\% slower
rate than in recent cycles.  Both of these facts indicate that cycle 24
is peculiar.  Analysis of the ``Rush to the Poles'' indicates that solar
maximum will occur in 2013 or 2014, but there is no indication of the
strength of the maximum.  There is at this time no confirmation of this
prediction from the southern hemisphere.

\acknowledgements 
The observations used herein are the result of a cooperative program of
the Air Force Research Laboratory and the National Solar Observatory.
I am grateful for the assistance of NSO personnel, especially John
Cornett, Timothy Henry, Lou Gilliam and Wayne Jones, for observing and
data-reduction and analysis services and maintenance of the Evans Solar
Facility and its instrumentation.

%%% THE BIBLIOGRAPHY 
%%% CONSULT SECTION 3 OF "INSTRUCTIONS FOR AUTHORS" FOR HOW TO USE NATBIB.  
%%% AUTHORS ARE ENCOURAGED TO USE EITHER THE "THEBIBLIOGRAPY" ENVIRONMENT 
%%% BY UNCOMMENTING (DELETING THE "%" SYMBOL) THE COMMANDS BELOW, OR BY 
%%% USING THE BIBTEX ENVIRONMENT. TO FIND OUT WHICH IS APPLICABLE TO YOUR 
%%% CONTRIBUTION, CONSULT THE VOLUME EDITORS FOR YOUR PROCEEDINGS.  

\end{document}